\documentclass[twocolumn,superscriptaddress,showpacs,amsmath,amssymb,aps,prl,floatfix]{revtex4}

\usepackage{graphicx}
\usepackage{dcolumn}
\usepackage{bm}

\usepackage[usenames,dvipsnames]{xcolor}
\usepackage{dsfont}
\usepackage{amsbsy}

\usepackage{amsfonts}
\usepackage{graphicx}
\usepackage{bbold}

\usepackage{calrsfs}
\DeclareMathAlphabet{\pazocal}{OMS}{zplm}{m}{n}

\makeatletter
\def\hlinewd#1{%
  \noalign{\ifnum0=`}\fi\hrule \@height #1 \futurelet
   \reserved@a\@xhline}
\makeatother

\begin{document}
\title{Decoherence of electron spins in isotopically enriched silicon near Clock Transitions}

\author{J.~E.~Lang }
\affiliation{Department of Physics and Astronomy, University College London,
Gower Street, London WC1E 6BT, United Kingdom}
\author{R.~ Guichard}
\affiliation{Department of Physics and Astronomy, University College London,
Gower Street, London WC1E 6BT, United Kingdom}
\author{S.~J.~ Balian}
\affiliation{Department of Physics and Astronomy, University College London,
Gower Street, London WC1E 6BT, United Kingdom}
\author{T.~S.~ Monteiro}
\affiliation{Department of Physics and Astronomy, University College London,
Gower Street, London WC1E 6BT, United Kingdom}
\date{\today}

\begin{abstract}
Despite the importance of isotopically purified samples in current experiments, there have been 
few corresponding studies of spin qubit decoherence using full
quantum bath calculations. Isotopic purification eliminates the well-studied nuclear spin baths which usually dominate decoherence.
 We model the coherence of electronic spin qubits in silicon near so called Clock Transitions (CT) where
 experiments have electronic $T_{2e}$ times of seconds. Despite the apparent simplicity of the residual decoherence mechanism, 
this regime is not well understood: the state mixing which underpins CTs allows also
  a proliferation of contributions from usually forbidden channels (direct flip-flops with non-resonant spins);
 in addition, the magnitude and effects of the corresponding Overhauser fields and other detunings is not well quantified.
For purely magnetic detunings, we identify a regime, potentially favourable for quantum computing, 
where forbidden channels are completely suppressed but
spins in resonant states are fully released from Overhauser fields and applied magnetic field gradients. 
We show by a general argument that the enhancement between this regime and the high field limit is $< 8$, regardless 
of density, while enhancements of order 50 are measured experimentally.
We propose that this discrepancy  is likely to arise from strains 
of exclusively non-magnetic origin, underlining the potential of CTs 
for isolating and probing different types of inhomogeneities. We also identify a set of fields,  ``Dipolar Refocusing Points" (DRPs), where the Hahn echo fully refocuses the effect of the dipolar interaction.

\end{abstract}





\maketitle
\section{Introduction}
There is considerable interest in the use of isotopically pure samples, with both diamond and silicon platforms, for implementation of solid state spin qubits for quantum information. Dramatic improvements
in performance have been measured in silicon \cite{Tyryshkin2012,Steger2012,Wolfowicz2013} with some of the longest coherence times ever observed.  Decoherence of the
electron qubit is dominated by  a bath 
of $^{29}$Si  nuclear spin impurities in natural silicon, or  $^{13}$C impurities in natural diamond, which lead to dephasing decoherence (spectral diffusion) of the electronic spin. The theory of decoherence by nuclear spin baths is well understood in both cases \cite{DeSousa, DeSousa2, Witzel1, Maze2008, Renbao2006, Zhao2012}. In purified samples, it is 
instead interactions between electronic donor spins themselves
which dominate decoherence in typical ESR experiments. For the case of silicon, many studies used ensembles, but there have also been a significant number of studies with single spin systems \cite{Morello2010, Pla2012, Pla2014}.

Although most implementations in silicon use phosphorus, there is also
 growing interest in other donor species \cite{George2010,Morley2010,Morley2013} such as bismuth, arsenic and antimony where there is
strong mixing between the donor electronic and nuclear spins at modest magnetic fields ($B_0 \lesssim 0.5$ T),
leading to rich and surprising coherence behavior. 
For these, Optimal Working Points (OWPs)  also referred to as CTs (``Clock Transitions'') 
\cite{OWP-CT} of enhanced electronic coherence have been identified,  and there have been several studies investigating these
\cite{Mohammady2010,Balian2012,Wolfowicz2013,Balian2014,Cywinski2014,Balian2015}.
However,  all the theoretical work on CTs/OWPs has to date been restricted to nuclear spin baths. 

\begin{figure}[htb]
\includegraphics[width=1.9in]{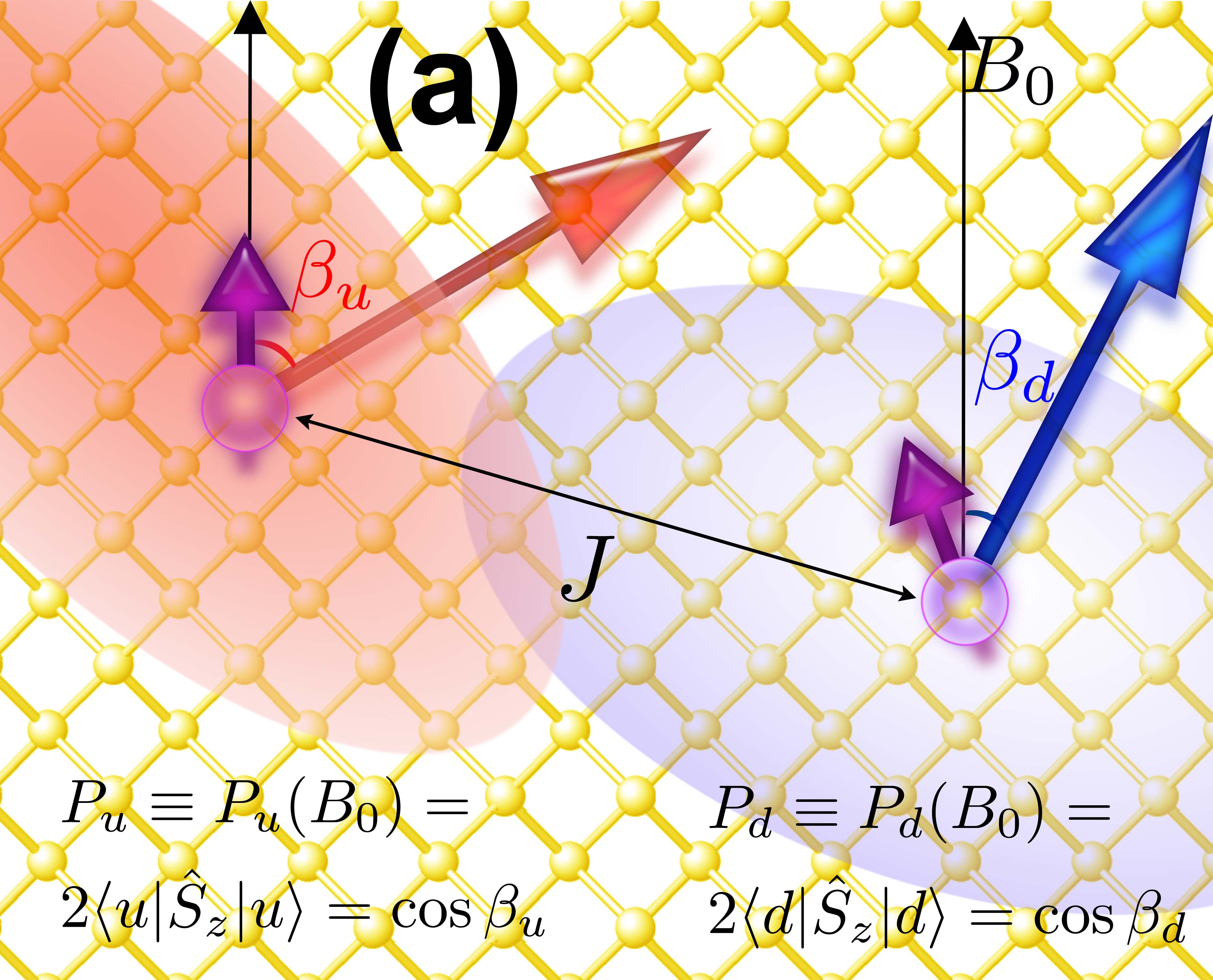}
\includegraphics[width=1.4in]{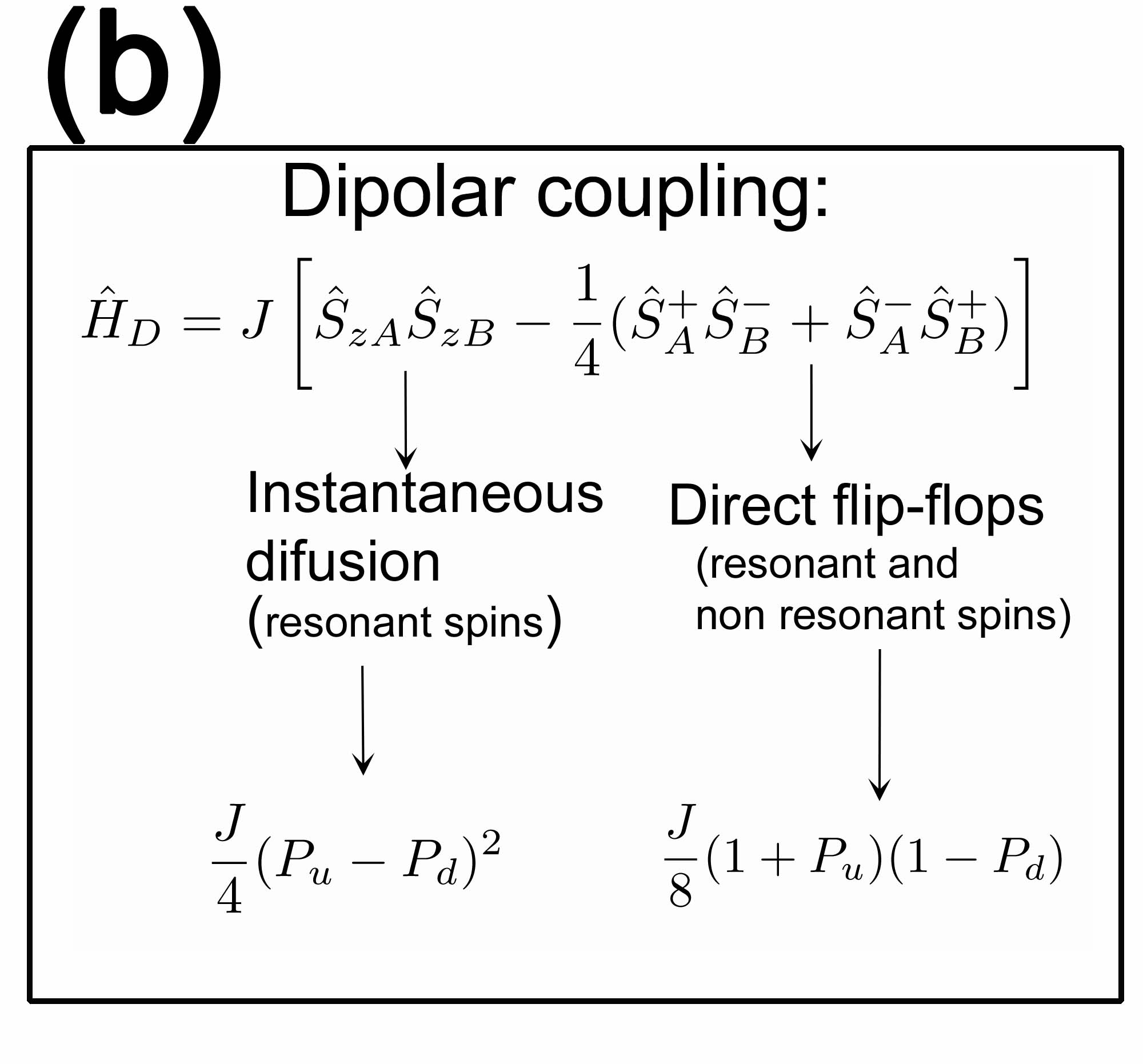}
\caption{ Illustrates two donor spin systems with dipolar interactions. {\bf (a)} The electronic spins have $s=1/2$, but
for atomic species with strong electronic coupling to the host nuclear spins (purple arrows) the  ${\hat S}_z$ eigenvalues $m_s=\pm 1/2$ are not good quantum numbers. Instead, each spin quantum state $i$ corresponds to eigenstates of an effective field, tilted to the $z$-axis by an angle $\beta_i$, with $P_i= \cos \beta_i= 2\langle i |{\hat S}_z|i\rangle$. Microwave pulses resonantly drive transitions $i= u \to d$ between two selected states; the figure exemplifies two donor atoms, each in one of these resonant spin states.
{\bf (b)}  The dipolar coupling allows two decoherence mechanisms: instantaneous diffusion (ID), a dephasing
arising in Hahn echo experiments where nearby spins are both rotated by the microwave pulse; and direct 
flip-flops; for ESR lines the relative magnitude of the contributions is given in terms of $P_u,P_d$.}
\label{Fig1}
\end{figure}

We  address this gap here by modelling the extremely long, measured 
$T_{2e}$ times in enriched samples, in particular a detailed experimental study of CT behavior
 for both  isotopically enriched bismuth (Si:Bi) and natural silicon \cite{Wolfowicz2013}. For
 Si:P there have been cluster correlation expansion (CCE) quantum simulations of the donor-donor spectral
 diffusion \cite{Witzel2012} (indirect flip-flop decoherence mechanism) which is the dominant mechanism away from CTs.
Near CTs, a quantum calculation of spectral diffusion becomes extremely challenging:
a many-body calculation (beyond pair correlation) is required to even 
obtain finite decay times \cite{Balian2014,Balian2015}. For the nuclear baths, good agreement with
 experimental results was obtained recently \cite{Balian2015} for Si:Bi, in natural silicon.

Fortunately, for enriched samples, spectral diffusion is suppressed at CTs and typically represents a negligible contribution.
The dynamics either (i) at the CT point $B_0 =B_\textrm{CT} $  or  at (ii) high fields $B_0  \gtrsim 0.5$T is comparatively simple.
(i) At the CTs, since spectral diffusion is strongly suppressed, the dominant process 
is the mutual decoherence of spins via direct flip-flops (DFF). 
(ii) At high fields there is a well-known
 dominant dephasing mechanism, instantaneous diffusion (ID) \cite{Raitsimring} (illustrated in Fig.\ref{Fig1}). ID is a dephasing
effect, arising because microwave  pulses, applied as part of the usual Hahn echo sequence
used to measure $T_2$ times, produce unwanted rotations of neighbouring resonant spins. Further details of
the spin mixing leading to CTs and of the decoherence dynamics are given in the Appendix.

\begin{figure}[hb]
\includegraphics[width=2.7in]{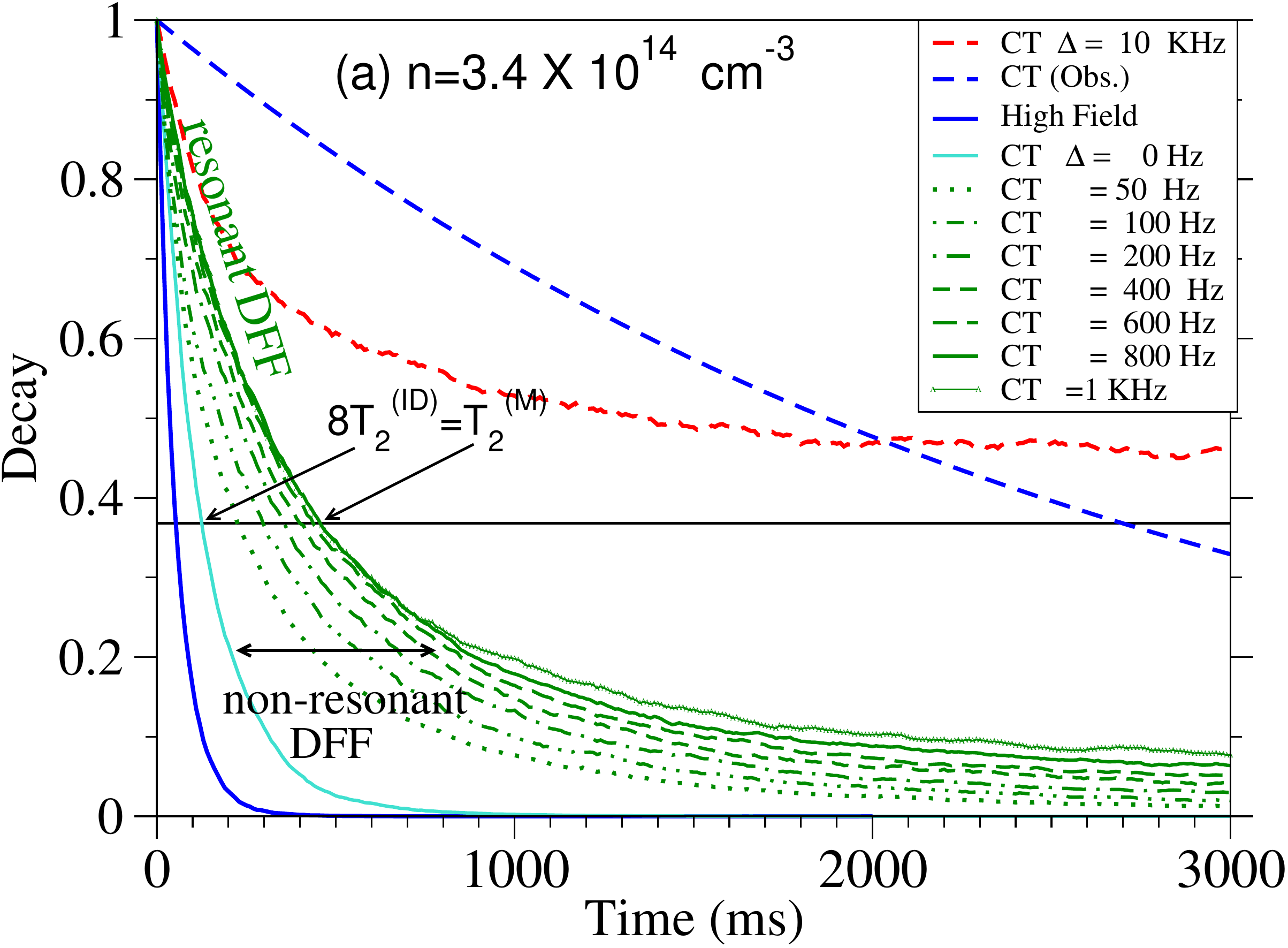}
\includegraphics[width=2.7in]{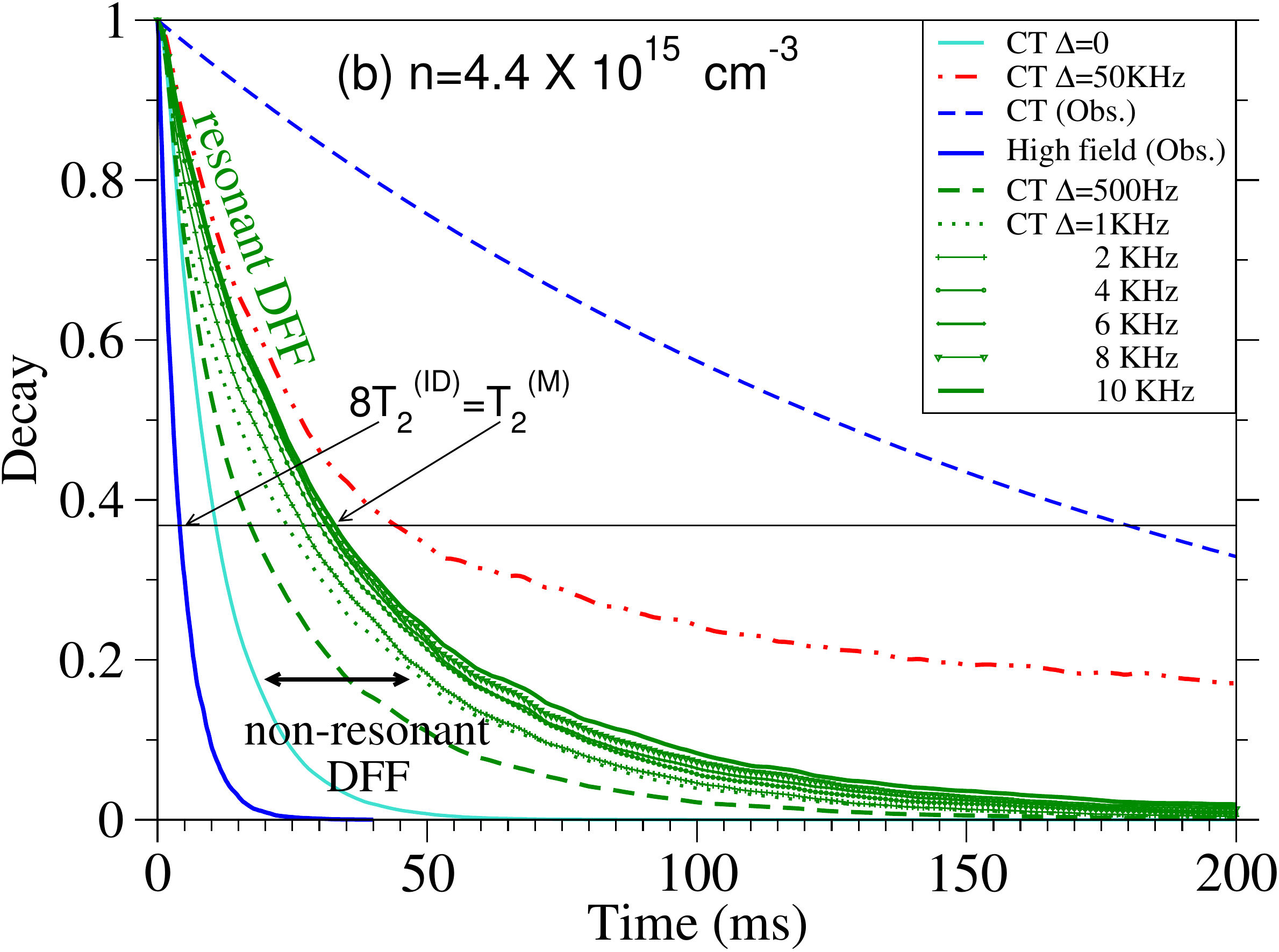}
\caption{Shows dependence of coherence decays (Hahn echo) on detuning fields of {\em magnetic origin} only  (Overhauser fields due to 
nuclear and electronic spins) which are Lorentzian-distributed, with width $\Delta=\gamma$
(see Eq.\ref{gamma}).
 The experimentally measured $14 \to 7$ transition of Si:Bi is investigated.
 At high magnetic fields $B_0 = B_\infty \gtrsim 1$ T (solid blue line) behavior is dominated by instantaneous diffusion
leading to a short coherence time $T_2^{(ID)}$.
At Clock Transitions (CT) direct flip-flops (DFF) are dominant (dashed blue line; 799G CT shown).
For $\Delta=0$, the clock transition offers very little enhancement over the high field limit,
because  of non-resonant flip-flops.
As $\Delta$ increases, non-resonant DFF decline;
the decays near a CT tend to a saturation value $T_2^{(M)}$, where
the resonant states are fully released from Overhauser fields, but  non-resonant channels are fully
suppressed. 
Here $T_2^{(M)}/T_2^{(ID)} \simeq [P_u(B_\infty)-P_d(B_\infty)]^2/[\frac{1}{2}(1+P_u(B_\text{CT}))(1-P_d(B_\text{CT}))]\simeq 8$.
 However, the observed $T_2$  are considerably longer; we conclude that
the discrepancy between $T_2^{(M)}$ and $T_2$ isolates inhomogeneities of non-magnetic origin,
 so may usefully probe different components of the detuning fields.
 {\bf (a)} low donor density  $n$ {\bf (b)} higher density $n$. The behaviors, to a good approximation, scale with $n$.
At typical Overhauser fields of $\sim 100$s Hz for purified samples with 50 ppm of $^{29}$Si, there is still an
appreciable contribution from non-resonant flip-flops.
At very large detunings the CT position itself becomes perturbed (dashed red line). }
\label{Fig2}
\end{figure} 

However, simulating the experimentally measured CT behavior in
enriched silicon  is still far from straightforward;
there is the richness introduced by the mixing: one must consider the proliferation of usually 
``forbidden'' channels allowing for non-resonant direct flip-flops (DFF). In
 systems like Si:P, with little state mixing, only flip-flops between resonant states (in
the sense of resonant with the applied microwave pulse) are considered.

Most importantly, the nature of the detuning fields, whether effective magnetic fields fields originating from other spins  or  contributions which are non-magnetic in origin
 (in the broad sense that they do not probe
 the $S_z$ component of the donor, which can be electric fields) is not well-understood.
The effective magnetic fields from both other donor spins as well as from residual $^{29}$Si 
are, for convenience, both grouped together and termed the Overhauser field (though commonly Overhauser fields
refer to the nuclear spins only).  These can be estimated numerically,
 given knowledge of the spin densities. But the non-magnetic inhomogeneities are not known. 

Nevertheless, even without precise quantitative knowledge of the detuning fields 
we can obtain some robust conclusions.
 We show that we can 
exploit the properties of the CT to isolate the effects  of different types of detunings on DFF and
propose a favourable operating regime for quantum information, where all forbidden channels are
 suppressed while resonant spins are fully released from the Overhauser field.
We show that in future qubit implementations, the fact that ID and DFF arise from the dipolar operator
means that they can interfere destructively in spin systems with mixing, leading to full refocusing
of the dipolar operator by a Hahn sequence, at arbitrary pulse spacing. The fields where this occurs are labeled ``Dipolar Refocusing Points" (DRPs).

\section{Theoretical model}
Instantaneous diffusion in experiments is usually analysed 
using a well-known approximate expression \cite{Raitsimring}, neglecting direct flip-flops.
 Yet both arise from the dipolar coupling 
between a single pair of spins which in its secular form is given by the Hamiltonian
${\hat H}_D= J \left [{\hat S}_{zA}{\hat S}_{zB} -\frac{1}{4}({\hat S}^+_{A}{\hat S}^-_{B}+{\hat S}^-_{A}{\hat S}^+_{B})\right] \equiv  H_{zz} +H_{ff}$  where $H_{zz}$ produces instantaneous diffusion while the flip-flop term
$H_{ff}$ limits coherence at CTs, since there $ H_{zz}=J{\hat S}_{zA}{\hat S}_{zB}$ energy shifts
between donors are strongly suppressed \cite{Mohammady2010}. The dipolar coupling strength is given by $J$.
 In the present work, we treat both terms as part of the same quantum bath process, allowing
interference between them as well as with the Overhauser fields. We define these by considering an
effective local magnetic field ${\mathcal{B}}_n$ felt by the $n-$th donor spin, generated by all other spins:
\begin{equation}
{\hat H}^{(n)}_{OH}= {\hat S}_{z n} \left[\sum_k J^{(n)}_k {\hat S}_{z}^{(k)} +
 \sum_{k'} A^{(n)}_{k'} {\hat I}_{z}^ {(k')}   \right] 
\equiv {\hat S}_{z n} {\mathcal{B}}_n
\label{Eq:OH}
\end{equation}
where the sum over $k$ represents the total field from other electron spins while
$k'$ denotes surrounding $^{29}$Si nuclei.
We can also include contributions from paramagnetic spin centers;
 we can calculate  ${\mathcal{B}}_n$ using each randomly generated configuration of spins.
These simulations indicate Lorentzian distributions for the calculated $\mathcal{B}_n$ distributions.
Thus in our simulations we obtain $\mathcal{B}_n$ from a randomly generated Cauchy distribution, characterised
by a half-width $\gamma$. To this we can also add some other detuning strain component of non-magnetic 
origin, $\delta_{NM}$ generated from an independently generated distribution.

We follow convention and consider one of the spins to be the qubit
spin of interest (``spin A'') and all other spins (whether resonant with the microwave pulse or not) 
to be the bath spins (`` B spins''). 
 The Hahn echo decay of spin A, $\mathcal{L}(t)= \langle S^+_A \rangle$ 
is  constructed from the product of all contributions from pairs formed with the $k$-th  bath spin: 
\begin{equation}
\langle\mathcal{L}(t)\rangle = \langle \prod_{k} {\mathcal{L}}_{k}(t)\rangle= \frac{1}{N}\sum_{j=1}^{j=N} \prod_{k} {\mathcal{L}}_{k}(t)
\label{clusters}
\end{equation}
where the sum denotes additional ensemble averaging over $N\simeq1000$ randomly generated configurations of donors  
with spin A at the center (each involving a product over pairs).
We consider first the case where both spins in the $k$-th pair are resonant with a microwave pulse.
 and the effect of the basic Hahn sequence $(\pi/2)_y - \tau -(\pi)_{x/y} - \tau-(\pi/2)_y$. 
  The effect of the first $(\pi/2)$ pulse on a pair in state
 $ |u\rangle_A |d\rangle_B$ or $ |u\rangle_A |u\rangle_B$ yield superpositions such as e.g.
 $ |u\rangle_A |d\rangle_B \to  \frac{1}{2}(|u\rangle_A+ |d\rangle_A) (|u\rangle_B- |d\rangle_B)$
while $ |u\rangle_A |u\rangle_B \to \frac{1}{2}(|u\rangle_A+ |d\rangle_A) (|u\rangle_B+ |d\rangle_B)$.
As shown in the appendix, atoms in the same state yield symmetric triplet states, atoms initially 
in different states yield singlet pair states.

The effect of the Overhauser Hamiltonian  ${\hat H}_{OH}$  is to introduce an energy cost $ \delta=E_{ud}-E_{du}$
between the states $ |u\rangle_A |d\rangle_B$ and $ |d\rangle_A |u\rangle_B$.
These are given in terms of the donor spin $z$-projections $P_u= 2\langle u |{\hat S}_z|u \rangle$
and $P_d= 2\langle d |{\hat S}_z|d\rangle$ as follows:
\begin{eqnarray}
\delta = \frac{1}{2}[(P_u\mathcal{B}_A + P_d\mathcal{B}_B)-(P_d\mathcal{B}_A + P_u\mathcal{B}_B)].
\end{eqnarray}
However, we can write  $ E_{ud}= \bar{E}+ \gamma/4$ and $E_{du}=\bar{E}- \gamma/4 $
where the mean  $\bar{E}= \frac{1}{4}(P_u + P_d)(\mathcal{B}_A+\mathcal{B}_B)$ is not dynamically significant.
The important splitting is:
\begin{equation}
\gamma= (P_u - P_d)(\mathcal{B}_A-\mathcal{B}_B).
\label{gamma}
\end{equation}

 It vanishes at the CT
as $P_u \to P_d$, regardless of the Overhauser magnetic splitting $\mathcal{B}_A-\mathcal{B}_B$, provided it is 
not large enough to perturb the mixing of the donor spin states (and hence the values of $P_u,P_d$). In our cluster numerics,
the $P_i$ are calculated always in the presence of the Overhauser field. But for typical Overhauser fields,
and even reasonable applied magnetic field gradients $\lesssim 10^3 \textrm{G/cm}$, the perturbation to the value of $P_i$
is negligible.

For such resonant spins we obtain,
\begin{equation}
 \mathcal{L}^\pm_{k}(t) = \frac{1}{2}(C^{+}_k e^{\pm i\frac{J_k }{4}(P_u-P_d)^2 t} + C^{-}_k e^{\mp i\frac{J_k }{4}(P_u-P_d)^2 t})
\label{TOTAL}
\end{equation}
where $C^{\pm}_k \equiv C^\pm_k(\Delta,J_k,P_u,P_d)$ are simple analytical expressions (derived in the Appendix).
The $+/-$ in the phases refer to whether the initial state was in a triplet/singlet superposition.
The $\Delta=\gamma+ \delta_{NM}$ where
$\delta_{NM}$ is the non-magnetic inhomogeneity.

\begin{figure}[htb]
\includegraphics[width=2.5in]{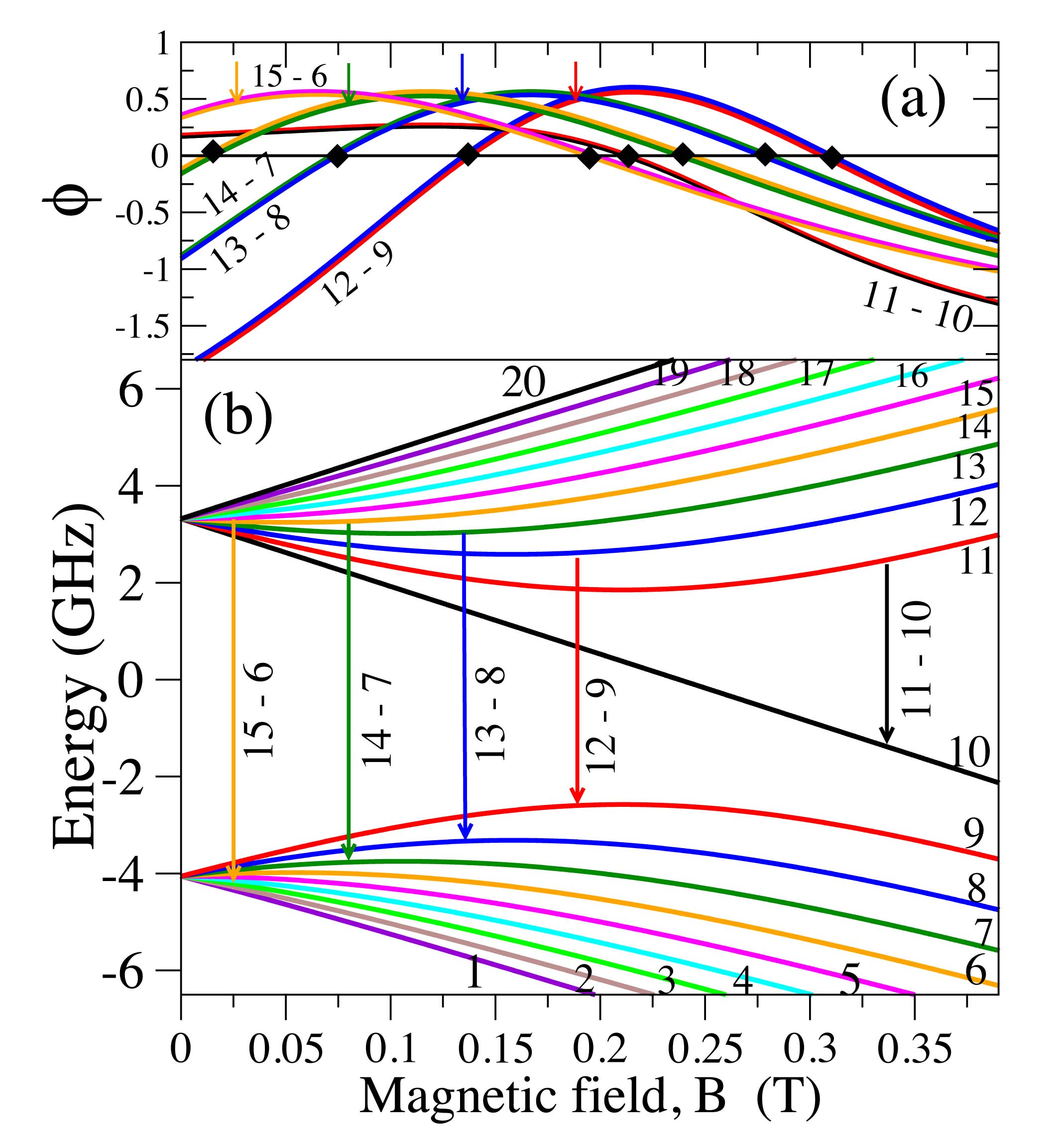}
\caption{ {\bf (a)} Position of Dipolar Refocusing Points (DRPs) for a few key ESR lines of bismuth.
 Figure plots  function $\phi= (P_u-P_d)^2- \frac{1}{2} (1+P_u)(1-P_d)=0$ in Eq.\ref{DRP1} and crossings of the $x$-axis (black diamonds) denote the DRPs. The two (or one) DRPs (which permit full suppression of the dipolar
interaction) for each transition
are at different field values to CTs where only diagonal, dephasing interactions are suppressed. The positions of
the CT for each transition are shown by arrows; the $14 \to 7$ CT has been studied experimentally in \cite{Wolfowicz2013}. {\bf (b) } Shows the energy level spectrum for bismuth. The transitions illustrated in (a) are indicated by arrows (at the corresponding OWP position, except for the 11-10 line which has a DRP but no CT). Doublets of constant $m$ are plotted in the same colour. }
\label{Fig3}
\end{figure}
For the case where one spin is not resonant, then, in the absence of detuning
\begin{equation}
 {\mathcal{L}}_{k}(t) = \cos{\frac{J_kt}{2}\rho}
\label{DFF}
\end{equation}
where $\rho = \langle u d |{\hat S}^+_{1}{\hat S}^-_{2}+{\hat S}^-_{1}{\hat S}^+_{2}| d u \rangle$.
In the presence of detuning fields, the expression is slightly  more complicated, but is given in the Appendix.

In the mixing regime ($B_0 \lesssim 0.5$T),
there are several such non-resonant transitions which are forbidden at high magnetic field $B_0$ which
lead to non-zero coupling via $S_1^+S_2^-$ terms at low fields due to the mixing.
For example , for the experimentally studied $14 \to 7$ CT at $799$G, there is some non-zero
probability of DFF at low fields if the neighbouring
spin B is in any of the states $i=5,6,8,12,13,15$ as well as $i=7,14$. The transition matrix elements for these are given in the Appendix.
For each of the $N$ configurations in Eq.\ref{clusters} the donors are randomly assigned to one of the $i=1-20$
spin states of the Si:Bi system and the appropriate contribution is calculated. Given the modest $B_0$ in
the mixing regime, polarisation effects are neglected.

\section{Results}

The results of the numerical cluster simulations and averages are shown in Fig.\ref{Fig2} for two densities
corresponding to the experiments in \cite{Wolfowicz2013} and for $\gamma$ ranging from $0 \to 50$ KHz.
This figure is a key result of this work.
One  striking feature shown at both low and high sample densities is a``saturation point'' where increasing the magnetic
detuning no longer causes an increase in the coherence time; we attribute this to the point at which all 
non-resonant flip-flop channels are fully suppressed. The other very evident result is the large difference
between this saturation point and the exponential decays which were measured at the CT, which corresponded
to a considerably longer $T_{2e}$ time.

To understand the results further, it is insightful to consider some limiting cases of Eq.\ref{TOTAL}. 
For  zero detuning, $\Delta=0$ for an ESR-allowed transition, Eq.\ref{TOTAL}  reduces to:
\begin{equation}
 {\mathcal{L}}^\pm_{k}(t) = \cos{\frac{J_k t}{4}\left[(P_u-P_d)^2\pm \frac{1}{2} (1+P_u)(1-P_d) \right]}
\label{IDDFF}
\end{equation}
Hence we can have  ${\mathcal{L}}_{k}(t) =1$ (thus zero decoherence) at magnetic
field values where:

\begin{equation} 
\phi= (P_u-P_d)^2- \frac{1}{2} (1+P_u)(1-P_d)=0,
\label{DRP1}
\end{equation}

 provided the spin pair were in a $ |u\rangle_A |d\rangle_B$ or $ |d\rangle_A |u\rangle_B$ and thus yielded a singlet state after the $\pi/2$ pulse. At these points, the ID and DFF effects interfere
destructively and the Hahn echo pulse fully refocuses the full dipolar interaction. 

Fig.\ref{Fig3} illustrates  a range of such dipolar refocusing points (DRPs) for Si:Bi. This refocusing effect is stronger than seen in a dynamical decoupling sequence as full recoupling occurs for arbitrary pulse interval $\tau$, in a Hahn sequence $(\pi/2)_y - \tau -(\pi)_{x/y} - \tau-(\pi/2)_y$;
but it requires that the spins in the  pair are prepared in specific states, thus the effect is not accessible with current ensemble experiments on Si:Bi but may become feasible in future applications with single spin pairs. Further details of DRPs are in the Appendix.

For the experimentally studied case,  at high fields, given detunings typical in silicon,
we have only the $H_{zz}$ contribution, thus we  obtain $C_k^{\pm}=1$ and:
\begin{equation}
 {\mathcal{L}}_{k}(t) = \cos{[\frac{J_k t}{4} (P_u-P_d)^2]}
\label{IDDFF}
\end{equation}
In this high-field limit, there is no DFF and the behaviour becomes fully insensitive to the Overhauser field. 

Conversely, near the CT, we see that  since $(P_u-P_d)(\mathcal{B}_A-\mathcal{B}_B) \to 0$ for the CT states
and they are free to flip-flop, while the ID term is eliminated,
hence, Eq.\ref{TOTAL} reduces to 
\begin{equation}
 {\mathcal{L}}_{k}(t) = \cos{[ \frac{J_k t}{8} (1+P_u)(1-P_d)]}
\label{DFF}
\end{equation}
for  a dipole allowed ESR line. For the CT in the experimental study \cite{Wolfowicz2013}, $P_u(B_\text{CT}) \simeq P_d(B_\text{CT})\simeq 0.1$;
in the high field (bare spin ) limit, $P_u(B_\infty)=+1,P_d(B_\infty)=-1$. 
In principle, the same spin pairs contribute to both limits, with a one-to-one mapping one can see the ID and DFF 
would involves a simple time rescaling of
$[P_u(B_\infty)-P_d(B_\infty)]^2/[\frac{1}{2}(1+P_u(B_\text{CT}))(1-P_d(B_\text{CT}))] \approx 8$. Thus, whatever the couplings distribution and
whatever the density or method of ensemble averaging, one might expect a factor $\sim 8$ increase in $T_2$ at the CT.
This is a significantly smaller enhancement than the factor $\sim 40-50$ measured. We show below that 
numerics using randomly generated pairs yield good agreement with the widely used ID expression
\cite{Raitsimring}. Thus, the discrepancy appears to arise entirely at the CT point and is fully independent of the form of the
Overhauser field (provided it is of magnetic origin and thus depends on $\hat{S}_{z}$).

\section{Discussion}

To a very good approximation, both measured decays as well as the coherence decays
calculated here, at either CTs and in the Instantaneous diffusion high-field
limit, were found to be of exponential form:
\begin{equation}
\langle\mathcal{L}(t)\rangle \simeq \exp(-t/T_{2e}). 
\label{formula}
\end{equation}
This is in contrast to spectral diffusion, where observed decays are typically $\langle\mathcal{L}(t)\rangle \simeq \exp{[(-t/T_{2e})^n]}$ with
$ n \approx 2$. Previous approaches to simulation of ID used stochastic approaches \cite{Raitsimring} which more naturally
give rise to exponentials; but here however, we use quantum bath methods to calculate the coherence,
with no random flip-flop rate introduced as in \cite{Raitsimring}. The one-cluster  forms $ \mathcal{L}_k(t)$ are not  dissimilar in form to those used for
spectral diffusion, thus it is important to understand how the exponential behavior arises.
In  \cite{Guichard2015} two types of clusters employed in modelling nuclear spin spectral diffusion gave very different
decays, although the $ \mathcal{L}_k(t)$ were of exactly the same form. The key difference was that in the one
case, the bath was numerous, thus the single central-spin decay is representative of all realisations. 
In other words the realisation average in Eq.\ref{clusters} (ie the sum)
$\langle\mathcal{L}(t)\rangle = \frac{1}{N}\sum_{j=1}^{j=N} \prod_{k} {\mathcal{L}}_{k}(t)$ 
is unimportant; only the cluster product is important. This behaviour typically gives
non-exponential decay. In contrast, for an alternative mechanism in \cite{Guichard2015}  
 only a few clusters contributed appreciably. Thus the behavior fluctuates considerably from one central spin to another;
this case yields decays with exponential character and extensive averaging $N\simeq 100-1000$'s is required to
obtain smooth decays.

For ID and undamped DFF, to a good approximation, the timescale of the decay is set by the single closest neighbouring 
donor which is resonant or can flip-flop. The more distant donors mainly serve to damp revivals of the oscillatory coherence
functions Eq.\ref{IDDFF} and Eq.\ref{DFF}. Thus we can provide a heuristic (but quite accurate) estimate of the coherence decay 
at the CT as the survival probability of each central spin state.

By using the binomial theorem, we can show that the 
 probability of a given donor having {\em no} neighbours closer than a distance $r$ is:
\begin{equation}
\mathcal{P}(r)=  \exp(-r^3/\overline{R}^3)
\label{exp}
\end{equation}
  where $ \overline{R}$ is a representative mean separation distance.

The dipolar coupling strength at a distance $r$, averaged over orientations is:
\begin{equation}
\langle{}J_k\rangle \simeq  \frac{1}{2}\frac{\mu_0}{4\pi}\gamma_e^2\hbar\frac{1}{r^3} = \frac{1}{2}\alpha\frac{1}{r^3}. 
\end{equation}
The characteristic timescale of decay, set by a neighbour at $r$ is found from equation Eq.~\ref{DFF} to be 
\begin{equation} \label{Eq: timescaleForR}
t_k \approx \frac{8}{\langle{}J_k\rangle(1 + P_u)(1 - P_d)} = \frac{16r^3}{\alpha(1 + P_u)(1 - P_d)}.
\end{equation}

The coherence is constructed as an average over many configurations of donor positions (Eq.~\ref{clusters}) and we can use the above 
to convert the exponential probability in Eq.\ref{exp} from $r$ into a temporal form $\mathcal{P}(r_k) \to \mathcal{P}(t_k)$. 
 If there is a nearest neighbour at distance $r$ then the coherence survives for a time $t_k$ as given above. 

We consider the regime where $T_2 = T_2^{(M)}$ thus where resonant spins flip-flop freely while all other channels are damped.
Taking then   $\frac{4\pi}{3} n_{res}  \overline{R}^3 \sim 1$ where
 $n_{res} \rightarrow n/10$ is the  density of resonant donor spins (only 10$\%$ of the spins are resonant with the central spin) we obtain the coherence decay at CT in this regime $\mathcal{L}_{CT}(t) \simeq \exp (-\frac{t}{T_{2}^{(M)} })$ with :
\begin{equation}
1/T_{2}^{(M)}  \approx \frac{\pi}{12}\frac{\mu_0}{4\pi}\gamma_e^2\hbar n_{res}
\end{equation}

For $n=4.4 \times 10^{21} \textrm{m}^{-3}$,  this yields $T_{2}^ {(M)} \simeq 37.5$ ms at the CT, in very good agreement with the cluster calculations, but
about a factor of $\sim 4$ shorter than the
measured value. Since $T_2^{(M)}/T_2^{(ID)} \simeq [P_u(B_\infty)-P_d(B_\infty)]^2/[\frac{1}{2}(1+P_u(B_\text{CT}))(1-P_d(B_\text{CT}))]$,
we also obtain $T_2^{(ID)} \simeq 4.5$ ms at high fields, at this density, in good agreement with standard expressions
for instantaneous diffusion.

An important question is whether highly enriched 50 ppm $^{29}$Si samples in \cite{Wolfowicz2013} 
are above or below the $T_2 = T_2^{(M)}$ regime or whether non-resonant spins contribute
apprecialy. A simulation for $n=4.4 \times 10^{21} \textrm{m}^{-3}$ indicates that the $^{29}$Si
dominate Overhauser fields with a Lorentzian $J$ distribution of width $\sim 6$KHz,
thus $\langle J\rangle P_{u,d}/4 \lesssim 200$ Hz, even including a contribution from
other donors. Fig.\ref{Fig2}(b) indicates that there is still a large, even dominant contribution 
from non-resonant spins.  
 Proximate $^{29}$Si nuclei (within tens of Angstroms
of the central spin) would provide a strong enough perturbation $ \gtrsim 10$ KHz to damp flip-flops
of resonant states at CTs; but for enriched samples, they will affect only a very small proportion of donors.

\section{Conclusion}

Given the modest Overhauser fields present in current experiments,
 we conclude that decoherence at Clock Transitions includes a very substantial contribution from
direct flip-flops with non-resonant spins.
  We have shown that quantum bath calculations cannot account for measured electronic
spin coherence $T_{2e}$ times at Clock Transitions
if only detuning fields of magnetic origin are included and so much of the enhancement
found over the high field limit is due to non-magnetic detunings of unknown origin.
Despite these uncertainties, some robust conclusions are still possible.
Were it not for non-magnetic detunings, the coherence times at CTs could be about
an order of magnitude shorter and enhancements over the high field limit
would be modest. The study identifies a regime where non-resonant flip-flops are fully
suppressed but where, because of the CT, resonant spins (not detuned by non-magnetic fields)
are fully released from the Overhauser field. Here they may interact and entangle via the dipolar
interaction, but may also be selectively detuned, if required,  by electric fields.
We identify also dipolar refocusing points where in future, if qubit detunings and
state preparation is possible, the full dipolar interaction can be eliminated by a simple Hahn
echo sequence of arbitrary pulse spacing.

{\em Acknowledgements}: We are grateful for helpful discussions and advice from  Gary Wolfowicz and Sougato Bose. JEL acknowledges an EPSRC DTA studentship. RG acknowledges funding from EPSRC grant EP/J010561/1.

\section{Appendix I: Donors and mixing}

The Hamiltonian of the donor spin system takes the form:
\begin{equation}
\hat{H}_0= \omega_0 \left( \hat{S}_z   - \delta  \hat{I}_z \right) +
A{\bf I} \cdot {\bf S},
\label{Eq:donor}
\end{equation}
where $\omega_0=B_0\gamma_e$ and $\gamma_e$ is the electronic gyromagnetic ratio, 
$\delta$ is the ratio of nuclear and electronic gyromagnetic ratios
and $A$ is the  isotropic hyperfine coupling.   For example, for 
Si:Bi,  $\frac{A}{2\pi}=1475.4 $~MHz while the $^{209}$Bi nuclear spin $I=9/2$
and $\delta = 2.488 \times 10^{-4}$.
In this case for magnetic fields $B \lesssim 0.3$T, we have the strong mixing regime
where $B_0\gamma_e= \omega_0 \sim A$. 
Then, the Zeeman quantum states $|m_s,m_I \rangle$ are not eigenstates of $\hat{H}_0$, but
the total $m=m_s+m_I$ is a good quantum number. 
We  consider magnetic fields where $A\gg \omega_I$, with  $\delta \ll 1$  and where the internuclear dipolar coupling is negligible thus we do not consider nuclear spin flips. We note also that the regimes we investigate are distinct from a recent studies of  phosphorus dimers 
which are close enough ($\sim6$ nm) to each other to be exchange-coupled and also have with $J \gg A$
\cite{dimers}; here we consider only dipolar coupled donors with $J \ll A$.

The form of the coupled electronic-nuclear spin eigenstates and eigenvalues in such regimes were given in refs.\cite{Mohammady2010}.
 There are $(2I+1)(2s+1) $ eigenstates,
ranging from 8 in total for arsenic to 20 states for bismuth for example. There are always two states for
which $|m|=s+I$ are aligned along the z-axis and remain unmixed. The other eigenstates  form doublets of constant $m$:
\begin{eqnarray}
&|+,m \rangle &=  \  \ \cos{\frac{\beta_m}{2}} |\frac{1}{2}, m-\frac{1}{2}\rangle + \sin{\frac{\beta_m}{2}} |-\frac{1}{2}, m+\frac{1}{2}\rangle \nonumber \\
&|-,m \rangle &=  -\sin{\frac{\beta_m}{2}} |\frac{1}{2},m-\frac{1}{2}\rangle+ \cos{\frac{\beta_m}{2}} |-\frac{1}{2}, m+\frac{1}{2}\rangle \nonumber\\
\label{mixed}
\end{eqnarray}
i.e, transformation from the Zeeman basis $|m_s, m_I\rangle$ to/from the eigenstate basis 
$|\pm,m \rangle$ are given by the rotation matrices  $R^T_y(\beta_m)$ and $R_y(\beta_m)$. 
Defining parameters $X_m=I(I+1)-m^2+1/4$ and $Z_m=m+\frac{\omega_0}{A}(1+\delta)$, then
the rotation angles  $\beta_m$ are given analytically by $\beta_m= \tan^{-1}X_m/Z_m$.

It can be seen from Eq.\ref{mixed} that
 $\langle \pm,m |{\hat S}_z| \pm,m\rangle= \frac{1}{2} \cos{\beta_m}=\frac{P_m}{2}$.
ESR transitions obey the selection rules $m_s-m'_s=\pm1, m_I-m'_I=0$ thus only  components of the 
$|\pm,m \rangle$ states which obey those same selection rules contribute to the line strength.
For transitions between states $u=|+,m\rangle \to d=|-,m-1\rangle$ 
which are dipole-allowed  at all fields,
we obtain:
\begin{eqnarray}
\rho &=& \langle u d |{\hat S}^+_{1}{\hat S}^-_{2}+{\hat S}^-_{1}{\hat S}^+_{2}| d u \rangle\nonumber \\
&=&  \cos^2{\frac{\beta_u}{2}} \cos^2{\frac{\beta_d}{2}} =\frac{1}{4}(1+P_u)(1+P_d)
\end{eqnarray}
In \cite{Mohammady2010} two types of forbidden transitions which have appreciable transition strengths at 
$B_0 \lesssim 0.3$T were identified; one class turns into NMR transitions at high-fields, the other is completely
forbidden as $B_0 \to \infty$. For the former, $\rho= \cos^2{\frac{\beta_u}{2}} \sin^2{\frac{\beta_d}{2}}$
while for the latter, $\rho= \sin^2{\frac{\beta_u}{2}} \sin^2{\frac{\beta_d}{2}}$;

\subsection{APPENDIX II: COHERENCE FUNCTION $\mathcal{L}(t)$}

We consider the evolution of the central donor (``spin A") coupled with the $k$-th bath donor spin (``spin B"). 
We assume that both spins in the $k$-th pair are resonant with a microwave pulse and analyse the effect of the basic Hahn sequence $(\pi/2)_y - \tau -(\pi)_{x/y} - \tau-(\pi/2)_y$. 
  The effect of the first $(\pi/2)$ pulse on a pair in state $ |u\rangle_A |u\rangle_B$ or $ |u\rangle_A |d\rangle_B$ yields the superpositions  $ |u\rangle_A |u\rangle_B \to \frac{1}{2}(|u\rangle_A+ |d\rangle_A) (|u\rangle_B+ |d\rangle_B)$ or $ |u\rangle_A |d\rangle_B \to  \frac{1}{2}(|u\rangle_A+ |d\rangle_A) (|u\rangle_B- |d\rangle_B)$.

The Hamiltonian for the spin A - spin B system is given by 
\begin{equation}  \label{AppEq:dFFHam}
\hat{H} = J\hat{S}_{Az}\hat{S}_{Bz} - \frac{J}{4}[\hat{S}^+_A\hat{S}^-_B + \hat{S}^-_A\hat{S}^+_B] 
 + \hat{S}_{Az}\mathcal{B}_A + \hat{S}_{Bz}\mathcal{B}_B
\end{equation}
where $J$ is the dipolar coupling strength and $\mathcal{B}_{A,B}$ are the Overhauser fields for each spin, as described in the main text.

After a $\pi/2$-pulse we have
\begin{equation}
|\psi(t=0)\rangle = \frac{1}{2} [ |u\rangle_A|u\rangle_B \pm |d\rangle_A|d\rangle_B ] 
                     + \frac{1}{2} [ |d\rangle_A|u\rangle_B \pm |u\rangle_A|d\rangle_B ]
\end{equation}
The $+/-$ factor refers to whether the spin B was initially in the state $|u\rangle_B / |d\rangle_B$. The system is allowed to evolve for a time $t = \tau$:
\begin{eqnarray}
& |\psi(t=\tau)\rangle=  \nonumber\\
	&\frac{1}{2} \Big[ \exp (-i[\frac{J}{4}P_u^2 + \frac{1}{2}P_u\mathcal{B}_A + \frac{1}{2}P_u\mathcal{B}_B]\tau)|u\rangle_A|u\rangle_B    \nonumber   \\ 
			  &  \pm \exp(-i[\frac{J}{4}P_d^2 + \frac{1}{2}P_d\mathcal{B}_A + \frac{1}{2}P_d\mathcal{B}_B]\tau)|d\rangle_A|d\rangle_B \Big]  \nonumber   \\  
               &   + \frac{1}{2} \Big[ a^\pm(\tau)|d\rangle_A|u\rangle_B \pm b^\pm(\tau)|u\rangle_A|d\rangle_B \Big].
\end{eqnarray}
The coefficients $a^\pm$ and $b^\pm$ are complex numbers that will be determined later.

Application of a $\pi$-pulse causes all the states to flip and then evolving for another time $\tau$ we obtain
\begin{eqnarray}   \label{AppEq: Psi2}
& |&\Psi(t=2\tau)\rangle = \frac{1}{2} e^{-i\phi\tau} \Big[ |d\rangle_A|d\rangle_B \pm |u\rangle_A|u\rangle_B \Big]   \nonumber  \\
 &                 & \ \ \ + \frac{1}{2} \Big[ a^\pm(2\tau)|u\rangle_A|d\rangle_B \pm b^\pm(2\tau)|d\rangle_A|u\rangle_B \Big]   \nonumber  \\
 &                 &= \frac{1}{2} |u\rangle_A \Big[ \pm e^{-i\phi\tau}|u\rangle_B + a^\pm(2\tau)|d\rangle_B\Big]  \nonumber  \\
&                  & \ \ \ + \frac{1}{2} |d\rangle_A \Big[ e^{-i\phi\tau}|d\rangle_B \pm b^\pm(2\tau)|u\rangle_B\Big] \nonumber \\
&                  &= \frac{1}{2} \Big[|u\rangle_A|\psi^\pm_u(2\tau)\rangle_B + |d\rangle_A|\psi^\pm_d(2\tau)\rangle_B \Big]
\end{eqnarray}
where the $\psi_{u,d}$ are the time evolved bath states after the Hahn echo sequence and   $\phi = \frac{J}{4}(P_u^2 + P_d^2) + \frac{1}{2}(P_u + P_d)(\mathcal{B}_A + \mathcal{B}_B)$.
The measured experimental signal is given by $\mathcal{L}_k(\tau) = \langle S_A^+\rangle$
To within a normalisation factor,
$\mathcal{L}_k(\tau) = \langle\psi^\pm_u(2\tau)|\psi^\pm_d(2\tau)\rangle$. This is given by
\begin{equation}
\mathcal{L}^\pm_k(\tau) 
			= (a^\pm(2\tau))^*   e^{-i\phi\tau}
			+  b^\pm(2\tau)  e^{+i\phi\tau}
\end{equation}
For typical ensemble experiments, the spins are not in pure states, but are in a thermal distribution of $u,d$ with equal probability as polarisation is weak at low magnetic fields. The measured signal is then an average over the four possible initial states 
of the pair $uu,dd,ud,du$,
so for the $k-$th pair, the measured signal is actually
\begin{equation}
tr(\rho_k S_A^+)= \langle \mathcal{L}^\pm_k(\tau)\rangle.
\label{bathav}
\end{equation}
 However, below we consider the behaviour of each 
component separately as we  allow for future preparation of qubits in pure states.

To determine the constants $a^\pm(2\tau)$ and $b^\pm(2\tau)$ we must consider the behaviour of the flip flopping part. In the basis $\left\{ |u\rangle_A|d\rangle_B, |d\rangle_A|u\rangle_B \right\}$  we can rewrite the Hamiltonian in Eq. \ref{AppEq:dFFHam} as
\begin{align}
&\tilde{H} =  \frac{J}{4} \times \nonumber \\
&
\begin{bmatrix}
& P_uP_d + 2(P_u\mathcal{B}_A + P_d\mathcal{B}_B)		& -\rho \\
& -\rho		& P_uP_d + 2(P_d\mathcal{B}_A + P_u\mathcal{B}_B)
\end{bmatrix}
\end{align}

where $\rho$ is defined in the previous section of the appendix.
We define $E_{ud} = \frac{1}{2}(P_u\mathcal{B}_A + P_d\mathcal{B}_B)$ and
 $E_{du} = \frac{1}{2}(P_d\mathcal{B}_A + P_u\mathcal{B}_B)$ and the mean $\bar{E} = \frac{1}{2}(E_{ud} + E_{du}) = \frac{1}{4}(P_u + P_d)(\mathcal{B}_A + \mathcal{B}_B)$. 
Then we can rewrite $E_{ud} = \bar{E} + \gamma/4$ and $E_{du} = \bar{E} - \gamma/4$. 
The quantity $\gamma = (P_u - P_d)(\mathcal{B}_A - \mathcal{B}_B)$ is the Overhauser detuning (after the
the dynamically unimportant average component $\bar{E}$ has been eliminated). Note that at the clock transition the Overhauser contribution vanishes. 

For atypically large detunings, we need to consider the situations where $P^{(A)}_{u,d} \neq P_{u,d}^{(B})$;
 In this case, the perturbed field $B_0+\mathcal{B}_{A/B}$ is used to evaluate $\langle S_z\rangle$.
In this case, $\bar{E} = \frac{1}{4}(P^{(A)}_u + P^{(A)}_d)\mathcal{B}_A + (P^{(B)}_u + P^{(B)}_d)\mathcal{B}_B$
and $\gamma = (P^{(A)}_u -P^{(A)}_d)\mathcal{B}_A + (P^{(B)}_u - P^{(B)}_d)\mathcal{B}_B$.
For convenience, we take $\mathcal{B}_A=0$ and take $\mathcal{B}_B$ from a Lorentzian distribution.
In this case, even when $(P^{(A)}_u -P^{(A)}_d)=0$ at the CT, one may find $(P^{(B)}_u -P^{(B)}_d) \neq 0$ 
and thus the Overhauser contribution does not vanish. 

With these quantities defined we can rewrite the above Hamiltonian as 
\begin{equation}
\tilde{H} = (\frac{J}{4}P_uP_d + \bar{E})\hat{\mathds{I}} +\frac{1}{4}
\begin{bmatrix}
&\gamma		& -J\rho \  \  \ \\
&-J\rho		& -\gamma
\end{bmatrix}
\end{equation}

Equation \ref{AppEq: Psi2} then becomes
\begin{align}
&|\psi(t=2\tau)\rangle  =   \nonumber \\
&\frac{1}{2} \exp(-i[\frac{J}{4}(P_u^2 + P_d^2) + 2\bar{E}]\tau) [ |d\rangle_A|d\rangle_B \pm |u\rangle_A|u\rangle_B ]    \nonumber\\ &+\frac{1}{2} \exp(-i[\frac{J}{4}P_uP_d +\bar{E}]2\tau) \times \nonumber\\
&[ a^\pm_\gamma(2\tau)|u\rangle_A|d\rangle_B \pm b^\pm_\gamma(2\tau)|d\rangle_A|u\rangle_B ]
\nonumber \\
& 
\end{align}
There is a global phase factor of $\exp(-i2\bar{E}\tau)$ which can be ignored. The coherence decay is then
\begin{equation}  \label{Eq:Overlap}
\mathcal{L}_k(\tau) 
			= (a^\pm_\gamma(2\tau))^*   e^{-i\frac{J}{4}(P_u - P_d)^2\tau}
			+  b^\pm_\gamma(2\tau)  e^{+i\frac{J}{4}(P_u - P_d)^2\tau}
\end{equation}

$a^\pm_\gamma(2\tau)$ and $b^\pm_\gamma(2\tau)$ are determined by calculating the evolution operator for the reduced Hamiltonian:
\begin{equation}
\tilde{\tilde{H}} =
\frac{1}{4}
\begin{bmatrix}
&\gamma		& -J\rho \  \  \ \\
&-J\rho		& -\gamma
\end{bmatrix}
\end{equation}
in the $\left\{ |u\rangle_A|d\rangle_B, |d\rangle_A|u\rangle_B \right\}$ basis. The eigenvalues of this Hamiltonian are $\omega = \pm \frac{1}{4} \sqrt{\gamma^2 + (J\rho)^2}$ and the eigenstates are tilted at an angle $\Theta = -\arctan(J\rho/\gamma) \equiv -\theta$. The time evolution operator is thus
\begin{equation}
\hat{U}_0(t) = \exp[-i\tilde{\tilde{H}}t] = \cos\omega{}t\hat{\mathds{I}} - i\sin\omega{}t\left[ -\sin\theta\hat{\sigma}_x + \cos\theta\hat{\sigma}_z \right]
\end{equation}
and the Hahn echo operator can be constructed as $\hat{U}(2\tau) = \hat{U}_0(\tau)\hat{\sigma}_x\hat{U}_0(\tau)$.
\begin{equation}
\hat{U}(2\tau) = A\hat{\sigma}_x + iB \hat{\mathds{I}} + C\hat{\sigma}_z
\end{equation}
where $A =\sin^2(\omega\tau)\cos{}2\theta + \cos^2(\omega\tau)$, $B = \sin(2\omega\tau) \sin\theta$ and $C = \sin^2(\omega\tau)\sin{}2\theta$. 
The effect of the Hahn echo is $\hat{U}(2\tau)[\pm |u\rangle_A|d\rangle_B + |d\rangle_A|u\rangle_B] = [a^\pm_\gamma|u\rangle_A|d\rangle_B \pm b^\pm_\gamma |d\rangle_A|u\rangle_B]$ which can be written as
\begin{equation}
\hat{U}(2\tau)
\begin{bmatrix}
 \pm & 1 \\ & 1
\end{bmatrix}
=
\begin{bmatrix}
 &iB + C & A\\ &A &iB - C
\end{bmatrix}
\begin{bmatrix}
 \pm & 1 \\ & 1
\end{bmatrix}
=
\begin{bmatrix}
&  a^\pm_\gamma  \\   \pm  &  b^\pm_\gamma
\end{bmatrix}
\end{equation}
This determines the values of $a^\pm_\gamma$ and $b^\pm_\gamma$ so using Eq.~\ref{Eq:Overlap} we obtain the result quoted in the main text 
\begin{equation}
\mathcal{L}^\pm_k(\tau) = \frac{1}{2}\Big[C^+_k e^{\pm i\frac{J}{4}(P_u - P_d)^2\tau} + C^-_k e^{\mp i\frac{J}{4}(P_u - P_d)^2\tau}\Big]
\end{equation}
where $C^\pm_k(\tau) = A \pm iB \mp C$

Near the clock transition the Overhauser detuning vanishes $\gamma \rightarrow 0$ which means $\theta = \pi/2$ and $\omega = J\rho/4$. Then $C = 0$ and $A \pm iB = \exp(\pm i2\omega\tau) = \exp(\pm iJ\rho\tau/2)$ so the coherence reduces to
\begin{equation}
\mathcal{L}^\pm_k(\tau) = \cos(J\rho\tau/2)
\end{equation}
In the high field limit the Overhauser detuning is large and supresses flip flops. This can be seen as $\theta \rightarrow 0$ , $B,C \rightarrow 0$ and $A \rightarrow 1$ so the coherence reduces to 
\begin{equation}
\mathcal{L}^\pm_k(\tau) = \cos(\frac{J}{4}(P_u - P_d)^2\tau).
\end{equation}

The bath-state averaged form of Eq.\ref{bathav} is especially simple:
\begin{equation}
\langle \mathcal{L}_k(\tau) \rangle =[\sin^2(\omega\tau)\cos{2\theta} + \cos^2(\omega\tau)]\cos(\frac{J}{4}(P_u - P_d)^2\tau)
\end{equation}
and reduces straightforwardly to the two limiting forms above when $\theta \to \pi/2 $ and $P_u \to P_d$ respectively.

For non resonant spins, in the presence of a detuning field,

\begin{equation}
\mathcal{L}_k(\tau) =1- \sin^2(\omega\tau)\sin^2{\theta}
\end{equation}
and this is the form used for the -usually dipole-forbidden at high-field additional channels where spin-B is
neither in state $u$ nor state $d$.

\section{Appendix III: Dipolar Refocusing Points} 

There is a set of field values 
where the full dipolar coupling can be eliminated, by a different mechanism involving the off-diagonal
interactions, for arbitrary times and coupling strengths, without the need for any complex dynamical decoupling sequences.
 A simple echo pulse suffices, since at these ``magic'' field values, which we refer to below as  dipolar refocusing points (DRPs), 
the donor spins' own internal level structure can lead to complete destructive interference
 between diagonal and off-diagonal dipolar contributions, with no requirement for pulse timing to be fast
compared with the internal dynamics.
 The effects of DRPs are not however  visible with thermal ensembles as the CTs and OWPs.
Thus the most practical applications may only become realisable when single spin or few spin techniques tested on phosphorus are extended to other donor species in silicon or if there are alternative advances which permit coherent control of donors.
We outline in brief their features.

We consider a spin system with eigenstates which are given by the donor spin Hamiltonian:
\begin{equation}
\hat{H}_0 |i\rangle = E_i  |i\rangle
\label{Eq:donor1}
\end{equation}

where $\hat{H}_0$ is given by Eq.\ref{Eq:donor}.
We consider qubits restricted to a  two-state space $i=u,d$ resonantly coupled by microwave radiation of frequency  $\hbar \omega_{u \to d}=E_u-E_d$.
Under the action of the Hamiltonians $\hat{H}_0+{\hat H}_D$, we use the usual two-spin triplet/singlet states,
$|T_{+1}\rangle=|u\rangle_A |u\rangle_B$,  $|T_{-1}\rangle=|d\rangle_A |d\rangle_B$, plus
$|T_{0}\rangle=\frac{1}{2} [|u\rangle_A |d\rangle_B + |d\rangle_A |u \rangle_B]$ and
$|S_{0}\rangle=\frac{1}{2} [|u\rangle_A |d\rangle_B - |d\rangle_A |u \rangle_B]$.

The resulting spectrum is illustrated in Fig.\ref{Fig4A}. We can show (see below) that
$\langle uu| {\hat H}_{ff}| uu\rangle = \langle dd| {\hat H}_{ff}| dd\rangle=0$ while
$\langle ud| {\hat H}_{ff}| du\rangle =\langle du | {\hat H}_{ff}| ud\rangle=\frac{J}{4} \rho$
 (analytical forms of $\rho$ as a function of $B_0$ for a given state $i=u,d$ are also given below).
Conversely, $\langle ud | {\hat H}_{zz}| du\rangle =\langle du | {\hat H}_{zz}| ud\rangle=0$ while
$\langle i j| {\hat H}_{zz}| i j\rangle=\frac{J}{4} P_i P_j  $ for any $i,j \equiv u,d$.
From the above, we can easily write down the time-evolved form of the eigenstates:
\begin{eqnarray}
|T_{+1}(t=0) \rangle \to |T_{+1}(t) \rangle= e^{-i(2E_u+ (J/4)P_u^2)t} |u\rangle_A |u\rangle_B\nonumber \\
|T_{-1 }(t=0) \rangle \to |T_{-1}(t) \rangle= e^{-i(2E_d+ (J/4)P_d^2)t}  | d\rangle_A |d\rangle_B
\end{eqnarray}
thus the $\pm 1$ triplet states do not lead to entanglement. 

However, if the qubits are
prepared in either the $|u\rangle_A |d\rangle_B $ or $ |d\rangle_A |u \rangle_B$ state, i.e
$|T_{0}\rangle \pm |S_{0}\rangle$ superpositions, the qubits become entangled, e.g.:
\begin{eqnarray}
|u\rangle_A |d\rangle_B \to \cos{(J\rho t)} \ |u\rangle_A |d\rangle_B +i \sin{(J \rho t)} \ |d\rangle_A |u\rangle_B
\label{Eq:tangle}
\end{eqnarray}
and likewise for  $|d\rangle_A |u\rangle_B$. It is this evolution which is eliminated by an echo sequence
at dipole refocusing points (DRPs) i.e. particular field values $B_0\equiv B_{DRP}$;
  more generally, at DRPs, the states 
$T_{+1}, T_{-1}, S_{0}$ become a decoherence-free subspace. 

This is easily seen by considering the
effect of the basic Hahn sequence $(\pi/2)_y - \tau -(\pi)_{x/y} - \tau-(\pi/2)_y$ on either
$ |u\rangle_A |d\rangle_B$ or $ |d\rangle_A |u\rangle_B$. The effect of the first $(\pi/2)_y$ pulse on
the former  is 
 $ |u\rangle_A |d\rangle_B \to  \frac{1}{2}(|u\rangle_A+ |d\rangle_A) (|u\rangle_B- |d\rangle_B)= 
\frac{1}{2}(|T_{+1}\rangle  - |T_{-1}\rangle+ \sqrt{2} |S_{0}\rangle)$ at $t=0$. This state evolves in time as follows:

\begin{eqnarray}
&2|\psi(t)\rangle{}&= e^{-i(2E_u+ \frac{J}{4}P_u^2)t} |u\rangle_A |u\rangle_B +
         e^{-i(2E_d+ \frac{J}{4}P_d^2)t}  | d\rangle_A |d\rangle_B \nonumber \\
       & + & e^{-i[(E_d+E_u)+ \frac{J}{4}(P_d P_u+\rho)]t} ( | u\rangle_A |d\rangle_B-|d\rangle_A |u\rangle_B)
\end{eqnarray}
Then, the $\pi$ pulse and subsequent evolution results in the state, at $t=2\tau$ (but for 
any $\tau$):

\begin{eqnarray}
&2|\psi(2\tau)\rangle{}&= e^{-i \frac{J}{4}(P_u^2+P_d^2)\tau}  (|u\rangle_A |u\rangle_B - | d\rangle_A |d\rangle_B) \nonumber  \\ 
                  &-& e^{-i \frac{J}{2}(P_d P_u+\rho)\tau}  (|u\rangle_A |d\rangle_B - | d\rangle_A |u\rangle_B 
\end{eqnarray}

where we disregard the inconsequential global phase $e^{-2i(E_u+E_d)\tau} $. We see that if 
$P^2_u+P^2_d=2(P_uP_d+\rho)$, we obtain
$\psi(2\tau)= \frac{1}{2}(|u\rangle_A- |d\rangle_A) (|u\rangle_B+ |d\rangle_B)$.
Then, the final $(\pi/2)_y$ pulse completely restores the initial state $ |u\rangle_A |d\rangle_B$.
Thus the entanglement of an initial state $ |u\rangle_A |d\rangle_B$ or $ |d\rangle_A |u\rangle_B$
can be fully controlled: if one (or repeated) Hahn pulses are applied, the effect of the dipolar interaction is eliminated.
If they are not, entanglement occurs as in Eq.\ref{Eq:tangle}.
We see also that since the $T_{+1}, T_{-1}$  states each acquire a phase $\phi_1=\frac{J}{4}(P_u^2+P_d^2)$
and the $S_0$ state acquires an identical phase $\phi_0= \frac{J}{2}(P_d P_u+\rho)=\phi_1$,
any arbitrary superposition of $T_{+1}, T_{-1}, S_{0}$ never dephases, regardless of $J$ or $\tau$.

The case for the state $T_{0}$ is different; this state has  phase 
$ \frac{J}{2}(P_d P_u-\rho)$ hence acquires a phase difference $=J\rho$ relative to the other three states.

For an allowed ESR transition, the DRP condition $P^2_u+P^2_d-2(P_uP_d+\rho)=(P_u-P_l)^2-2\rho=0$ becomes:

\begin{equation}
 \phi= (\cos{\beta_u}-\cos{\beta_d})^2- 2\cos^2{\frac{\beta_u}{2}} \cos^2{\frac{\beta_d}{2}}=0
\label{DRP}
\end{equation}
and to find the DRP points, we can search for solutions of the above as a function of magnetic field $\omega_0$.
For forbidden transitions, $\rho= \cos^2{\frac{\beta_u}{2}} \sin^2{\frac{\beta_d}{2}}$
while for the latter, $\rho= \sin^2{\frac{\beta_u}{2}} \sin^2{\frac{\beta_d}{2}}$; in these cases, the
form of Eq.\label{DRP} must be adjusted, accordingly, to find the corresponding DRPs.

\begin{figure}[h]
\includegraphics[width=2.8in]{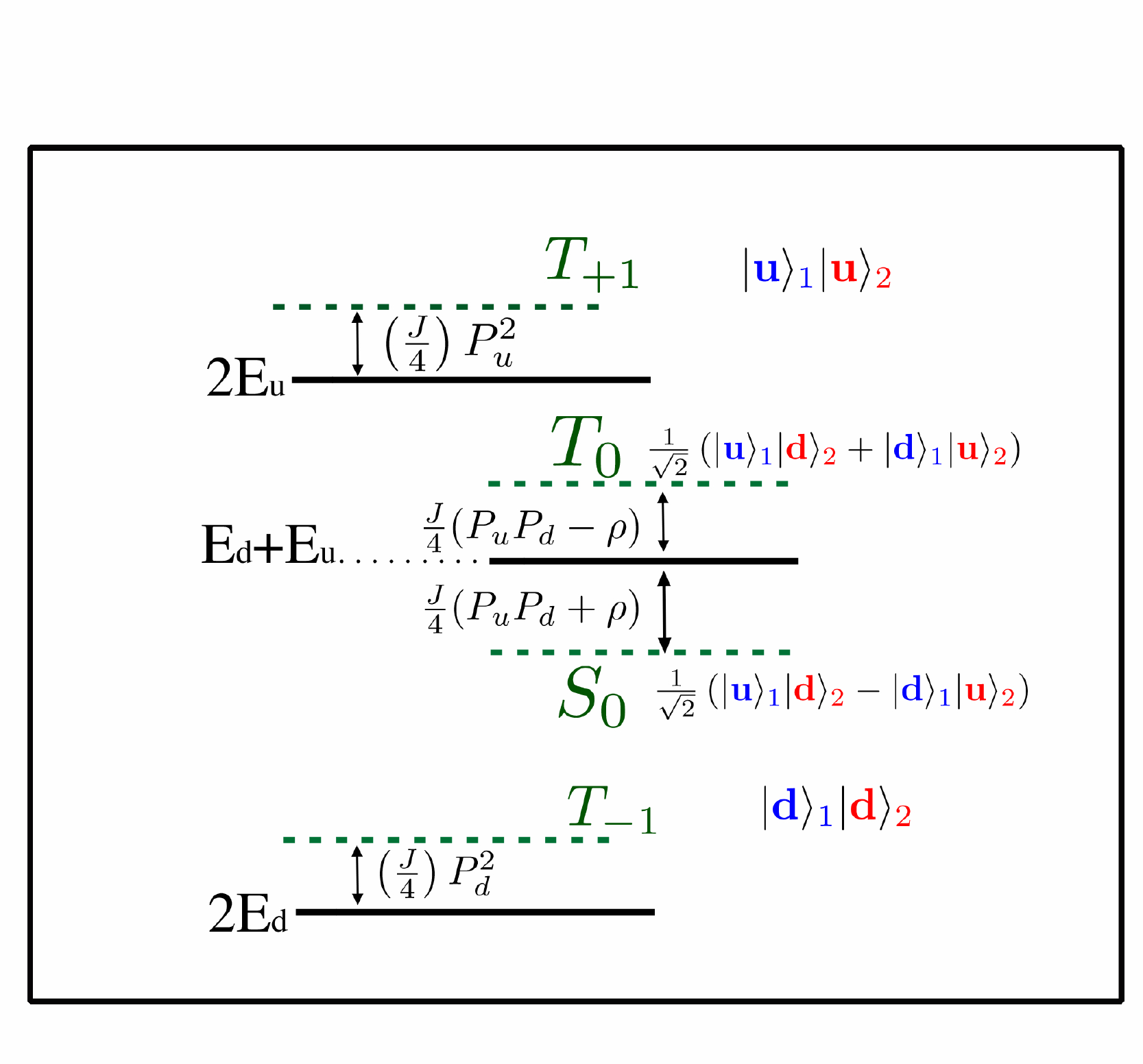}
\caption{ Represents the energy levels of an interacting pair of spin systems. 
At high magnetic fields, without mixing, the states of the $u \to d$ transition yield a triplet of states ($T_{+1},T_0, T_{-1}$) and a singlet $S_0$. With mixing,
a similar structure is preserved, but the energy shifts due to $H_D$ (shown in the figure) are 
field-dependent and are given by 
$P_u,P_d$ and $\rho=\langle u d |{\hat S}^+_{1}{\hat S}^-_{2}+{\hat S}^-_{1}{\hat S}^+_{2}| d u \rangle$.
DRPs occur at field values where $P^2_u+P^2_d=2(P_uP_d+\rho)$. }
\label{Fig4A}
\end{figure}


\begin{thebibliography}{99}
\bibitem{OWP-CT} Here, OWP (Optimal Working Point) denotes  a field point where there is strongly enhanced coherence; while a CT (Clock Transition) refers to a magnetic field point where
$df/dB=0$. While they are often near-coincident, as shown previously, not all CTs are OWPs; and in the present work we identify OWPs which are not at  CTs.
\end{thebibliography}
\end{document}